\documentstyle[aps]{revtex}

\begin{document}
\draft
\preprint{}
\title{    
Impurity effects on the spin excitation spectra in a d-wave superconductor}
\author{
Jian-Xin Li, Wei-Guo Yin}
\address{
Department of Physics, Nanjing University, Nanjing 210093, People's Republic
of China}
\author{
Chang-De Gong}
\address{
Chinese Center of Advanced Science and Technology (World Laboratory), P.O. Box 8730,\\ 
Beijing 100080, People's Republic of China\\
and National Laboratory of Solid State Microstructure, Nanjing University,
Nanjing 210093, People's Republic of China}
\maketitle

\begin{abstract}
The effects of nonmagnetic impurity on the spin excitation spectra in a 
$d_{x^{2}-y^{2}}$-wave superconductor are examined, using the self-consistent $t$-matrix approximation. It is shown that the impurity self-energy acts to shift the position of the resonance peak to low frequencies and broaden the peak. While the impurity vertex correction causes a broad spectral weight in the spin gap at the impurity concentrations where no clear resonance peak is observed. The gaplike feature still remains in low frequency region upon the introduction of impurities. Incorporating these two effects, we find that the result is in qualitative agreement with experiments
on YaBa$_{2}$(Cu$_{1-x}$Zn$_{x}$)$_{3}$O$_{6+y}$.
\end{abstract}


           
\section{INTRODUCTION}
  The spin excitation spectra Im$\chi$ of high-$T_{c}$ superconductors is 
extensively  studied by inelastic neutron scattering(INS) and a consistent 
picture has emerged in 
YBa$_{2}$Cu$_{3}$O$_{6+x}$(YBCO). A remarkable feature in both underdoped and
highly doped YBCO is that a sharp neutron resonance peak was observed in the 
superconducting state(SC) at a 2D wave vector {\bf Q}=$(\pi,\pi)$[1-4]. Also, 
in the SC state, Im$\chi$ is restricted to a small energy range 
limited by a doping dependent energy gap(spin gap) in low frequencies and
has sinusoidal dependence on $q_{z}$, the wave vector perpendicular to the 
CuO$_{2}$ planes~\cite{mook,fong}.
Both the resonance peak and the spin gap disappear in the normal state,
and the resonance energy E$_{r}$ is found to increases monotonically 
with the superconducting transition temperature $T_{c}$~\cite{fong,reg}, 
therefore they appear to be correlated to the superconductivity. 
  
  A number of theories has been proposed to account for this magnetic 
resonance. Beyond different modifications, one may basically divide the 
explanations
into two classes. First, it may result from the spin-flip quasiparticle
scatterings across the SC gap, causing an enhancement of the electronic
spin susceptibility at a specific energy which compensates for the loss
of spectral weight below the gap. Second, it may
be a consequence of a collective mode in the particle-particle channel 
which couples to neutron scattering through the particle-hole($p-h$) mixing 
in the d-wave SC state~\cite{demler}. More particularly, the first class 
includes i) a BCS gap function wth
strong Coulomb correlations~\cite{orig,blu,bulut} and a non-BCS gap function
resulting from the interlayer pair tunneling theory of high-$T_{c}$ 
superconductivity~\cite{yin}, in the framework of a d-wave pairing model.
ii) a s-wave order parameter with opposite signs in bounding and 
antibounding bands formed within the CuO$_{2}$ bilayer~\cite{mazin}. 

Experimentally, it was shown that the superconducting properties are 
modified by nonmagnetic impurities, especially $T_{c}$ is
rapidly suppressed by the substitution of the copper ions by zinc 
ions~\cite{xiao}. A possible interpretation of these results is made in
term of a $d_{x^{2}-y^{2}}$ order parameter affected by nonmagnetic 
scattering, assuming that Zn acts as a strong resonant scatterer
~\cite{hot,sun,hir1,feh}. In this case, nonmagnetic impurities have a 
strong pair breaking effect~\cite{hir1,feh}
and will modify the spin excitation spectra observed in SC state.

The purpose of this paper is to study the modifications of the resonance 
peak and the spin gap upon the introduction of nonmagnetic impurites in 
a BCS $d_{x^{2}-y^{2}}$ superconductor. We treat the impurities in the 
dilute limit using the self-consistent $t$-matrix approximation~\cite{hir2}.
Both the impurity self-energy effects and the impurity vertex corrections 
are considered in our calculations. In the pure system, a sharp magnetic
peak is observed and the spectrum is limited by the spin gap in low
frequencies. When the impurity self-energy corrections are considered, we
find that the peak is broadened and its position is shifted to lower energies.
Meanwhile, the magnitude of the spin gap decreases but only a negligible
contribution to the spin excitation spectrum is found in the gap due to the
impurity self-energy corrections. On the other hand, the vertex corrections
alone induce a broad spectral weight in
the spin gap at the impurity concentration where no clear reasonance peak
is observed, and have only slight effect on both the peak and the magnitude
of the spin gap. 

The paper is organized as follows. In Sec.II, we discuss the model and 
study the self-energy corrections. Sec.III contains the effect of the
vertex corrections. We present the conclusion in Sec.IV

\section{THE SELF-ENERGY CORRECTION}

To consider the modulation of the spin susceptibility along the $c$ axis, we
investigate a bilayer system with a interlayer hopping $t_{\perp}$ and the 
same $d_{x^{2}-y^{2}}$ order parameter in two layers. The effects of 
antiferromagnetic correlations within and between the layers are considered 
in the 
random-phase approximation(RPA) form of the susceptibility. The Nambu 
Green's function of single particles for the pure system in the SC state is  
given by,
\begin{equation}
\hat{g}_{0}^{(i)}({\bf k},i\omega_{n})={i\omega_{n}\hat{\sigma}_{0}
+\Delta_{k}\hat{\sigma}_{1}+\xi_{k}^{(i)}\hat{\sigma}_{3} \over
(i\omega_{n})^{2}-\Delta_{k}^{2}-(\xi_{k}^{(i)})^{2}} ,
\end{equation}
where $\hat{\sigma}_{i} (\hat{\sigma_{0}}=\hat{\bf 1})$ is the Pauli
matrices, $i=a$ or $b$ expresses the antibonding or bonding band. For the
quasiparticle dispersion, we use $\xi_{k}^{(a/b)}=-2t(\cos k_{x}+\cos k_{y})
-4t^{'}\cos k_{x}\cos k_{y}-2t^{''}[\cos(2k_{x})+\cos(2k_{y})] \pm t_{\perp}
-\mu$, with $t^{'}/t=-0.2, t^{''}/t=0.25, t_{\perp}/t=0.44, \mu/t=-1.11$
corresponding to a fit to the angle-resolved photoemission data on the 
optimal doping YBCO as used before~\cite{blu}. The order parameter is chosen 
as $\Delta_{\bf k}=\Delta_{0}(\cos k_{x}-\cos k_{y})/2$, where 
$\Delta_{0}=4T_{c0}$ and $T_{c0}$ the SC transition temperature.

The nonmagnetic impurities are modeled by a zero-range potential $V$ and its
scattering is treated in the self-consistent $t$-matrix approximation~\cite{hir2}.
In this approach, two parameters are introduced to describe the scattering 
process: $c=1/(\pi N_{0}V)$ and $\Gamma =n_{i}/\pi N_{0}$, where $N_{0}$ and 
$n_{i}$ are respectively the density of states at the Fermi level and the
impurity concentration. 
The impurity-average Nambu Green's function $\hat{g}({\bf k},i\omega_{m})$
for single particles can be written formally as~\cite{hir1,hir2},
\begin{equation}
\hat{g}^{(i)}({\bf k},i\omega_{n})={i\tilde{\omega}^{(i)}_{n}\hat{\sigma}_{0}
+\Delta_{k}\hat{\sigma_{1}}+\tilde{\xi}_{k}^{(i)}\hat{\sigma}_{3} \over
(i\tilde{\omega}^{(i)}_{n})^{2}-\Delta_{k}^{2}-(\tilde{\xi}_{k}^{(i)})^{2}} .
\end{equation}
The tilde symbol represents inclusion of the impurity self-energy corrections,
\begin{equation}
\tilde{\omega}_n^{(i)}=\omega_{n} -\Sigma^{(i)}_{0}(\omega_{n}) , \
\tilde{\xi}^{(i)}_{p}=\xi_{p}^{(i)}+\Sigma_{3}^{(i)}(\omega_{n}) ,
\end{equation}
where we have used the fact that the off-diagonal self-energy $\Sigma^{(i)}_{1}$
vanishes for a $d_{x^{2}-y^{2}}$ symmetry of the gap function.
In the single-site approximation, the self-energy is given by
$\Sigma^{(i)}_{j}=\Gamma T_{j}^{(i)}$. The impurity-scattering $t$-matrix
$T_{j}^{(i)}$ can be calculated from~\cite{hir1,hir2},
\begin{equation}
T_{0}^{(i)}={G^{(i)}_{0}(\omega) \over c^{2}-[G^{(i)}_{0}(\omega)]^{2}} , \
T_{3}^{(i)}=-{c \over c^{2}-[G^{(i)}_{0}(\omega)]^{2}} ,
\end{equation}
with $G_{0}^{(i)}(\omega)=(1/\pi N_{0})\sum_{k}{\rm Tr}[\hat{\sigma}_{0}
\hat{g}^{(i)}({\bf k},\omega)]$.
The following calculations are carried out in the unitary
limit, $c=0$, so only the $\Sigma_{0}$ contribution remains. The order parameter
$\Delta (\Gamma,0)$ and the SC transition temperature $T_{c}$ in the presence
of impurities are determined from the gap equation. In the weak-coupling
limit, it has been shown that $\Delta (\Gamma,0)/\Delta_{0}$ and $T_{c}/T_{c0}$
draw almost the same curve as a function of $\Gamma$, i.e.,
$\Delta (\Gamma,0)/\Delta_{0}\approx T_{c}/T_{c0}$~\cite{sun}. The
temperature-dependence of $\Delta (\Gamma,T)$ is taken to be,
\begin{equation}
\Delta (\Gamma,T)=\Delta (\Gamma,0)\tanh (2\sqrt{(T_{c}/T)-1}) ,
\end{equation}
where $T_{c}$ is given by the Abrikosov-Gor'kov formula~\cite{abr},
\begin{equation}
-\ln ({T_{c} \over T_{c0}})=\psi ({1\over 2}+{\Gamma \over 2\pi T_{c}})
-\psi ({1\over 2}) ,
\end{equation}
with $\psi(x)$ the digamma function.

The spin susceptibility for Matsubara frequencies is calculated from,
\begin{equation}
\chi^{(ij)}_{0}({\bf q},i\omega_{m})=-T\sum_{n}\sum_{k}{\rm Tr}
[{1\over 2}
\hat{g}^{(i)}({\bf k},i\omega_{n})\hat{g}^{(j)}({\bf k+q}, 
i\omega_{m}+i\omega_{n})] .
\end{equation}
Its analytic continuation to the real frequency, giving $\chi_{0}^{(ij)}
({\bf q},\omega)$, is performed using Pad$\acute{e}$ approximants~\cite{vid}.
When $\hat{g}^{(i)}$ is replaced by $\hat{g}_{0}^{(i)}$, Eq.(7) gives the
result for the pure system. The antiferromagnetic correlations in the plane
$J_{\Vert}$ and between the planes $J_{\perp}$ would renormalize
$\chi^{(ij)}_{0}$. This effect is considered in RPA approximation~\cite{ubben},
\begin{equation}
\chi^{(ij)}({\bf q},\omega)={\chi^{(ij)}_{0}({\bf q},\omega) \over
1+J^{+}({\bf q})\chi^{(ij)}_{0}({\bf q},\omega)} ,
\end{equation}
with $J^{+}({\bf q})=J({\bf q})-J_{\perp}$ and $J({\bf q})=J_{\Vert}(\cos q_{x}
+\cos q_{y})$.
We note that the susceptibility described above comes from the $p-h$
excitations of quasiparticles within and between the bonding and antibonding
bands. However, the susceptibility $\chi^{ph}({\bf q},\omega)$ observed 
in the neutron scattering is related to the excitations of 
quasiparticles within and between the layers~\cite{sato}. 
The relation between them can be obtained using the transformation matrix 
between the states in the layer and band representations. It gives,
\begin{equation}
\chi^{(11)}=\chi^{(22)}={1\over 4}[\chi^{+}+\chi^{-}] , \ 
\chi^{(12)}=\chi^{(21)*}={e^{-iq_{z}d}\over 4}[\chi^{+}-\chi^{-}] , 
\end{equation}
where $d$ is the distance between two layers, $\chi^{+}
=\chi^{(aa)}({\bf q},\omega)+\chi^{(bb)}({\bf q},\omega)$ and 
$\chi^{-}=\chi^{(ab)}({\bf q},\omega)+\chi^{(ba)}({\bf q},\omega)$.  
Then we have,
\begin{equation}
\chi^{ph}({\bf q},\omega)=\chi^{(11)}+\chi^{(12)}
+\chi^{(21)}+\chi^{(22)}= \chi^{+} \cos^{2}{q_{z}d \over 2}
+\chi^{-} \sin^{2}{q_{z}d \over 2} .
\end{equation}
Eq.(10) implies that the experimentally observed $\sin^{2}(q_{z}d/2)$ modulation 
of the INS comes from the transitions of quasiparticles between the 
respective bands. 

In following evaluations, the summation over ${\bf k}$ and $n$ are performed by
dividing the Brillouin zone into 1024$\times$1024 lattices and by summing 
from $n=-100$ to $n=100$ in Matsubara frequency $\omega_{n}=\pi T(2n-1)$, 
respectively. The number of input points in Pad$\acute{e}$ approximant is 
chosen to be 100 and $J^{+}({\bf Q})$ to be 0.85 in unit of $t$(We will use
this unit in the following). In addition, we take $T_{c0}=0.1$ and $T=0.1T_{c0}$.

Results for Im$\chi^{ph}({\bf Q},\omega)$ versus $\omega$ are shown in Fig.1. 
The continuous line corresponds to the pure system which reproduces the 
observed INS features in the SC state. The dashed (dotted) lines are results
with the impurity self-energy corrections. To understand the impurity effect, 
let us first address the origin
of the peak for the pure system within the $d$-wave BCS framework, which has
been studied in Ref.~\cite{blu}. For a qualitative statement, let $T=0$, 
and set the coherence factor to unity, then one has Im$\chi_{0}^{(ij)}
({\bf Q},\omega)=\pi \sum_{k} \delta(\omega-E_{k}^{(i)}-E_{k+Q}^{(j)})$.
The energy $E^{(ij)}({\bf k})=E_{k}^{(i)}+E_{k+Q}^{(j)}$ which is the 
function of the 2D wave vector ${\bf k}$ has a minimum at $E_{min}^{(ij)}
({\bf k})\approx 2\Delta_{0}=0.8$, corresponding to both $\bf k$ and 
$\bf k+Q$ near the crossings of the Fermi surface and the magnetic 
Brillouin zone. At the minimum Im$\chi_{0}^{(ij)}({\bf Q},\omega)$
has a step and correspondingly Re$\chi_{0}^{(ij)}({\bf Q},\omega)$ has a
logarithmic singularity. In the realistic calculations, this divergence
exhibits a maximum as shown in Fig.2 and causes a resonant peak due to
the RPA renormalization Eq.(8). Meanwhile, there is a saddle point
at $(0,\pi)$ in the quasiparticle dispersion, and it leads to a logarithmic
divergence in Im$\chi_{0}^{(ij)}({\bf Q},\omega)$. It arises from the transitions 
between the
occupied states located at $(0,\pi)$ and empty states above the SC gap, thus
the peak position locates at $E_{sp}^{(i)}=\Delta_{0}+\sqrt{\Delta_{0}^{2}
+(\xi_{vH}^{(i)})^{2}}$. For the dispersion of quasiparticles considered here, the
van Hove singularity of the antibonding band at $(0,\pi)$ lies at an energy
$\xi_{vH}^{(a)}=-0.25$ relative to the Fermi level and that of the bonding band is
$\xi_{vH}^{(b)}=-1.12$ due to splitting of the two bands. It gives $E_{sp}^{(a)}\approx
0.87$, which is close to the energy where Re$\chi_{0}^{(ij)}({\bf q},\omega)$
is divergent,
therefore enhances the peak. In fact, these two effects are indistinguishable
in the calculations and exhibits only one peak in Im$\chi_{0}^{(ij)}({\bf Q},\omega)$ as can
be seen in Fig.2. Now, we turn to the impurity effects on the spin excitation
spectra. In a $d$-wave superconductor with resonant nonmagnetic impurity
scattering, the SC gap $\Delta_{0}$ is suppressed~\cite{hot,sun} and causes 
the shift of the peak, which is basically 
equal to $2\Delta_{0}$ as discussed above, to low frequencies. 
Meanwhile, the impurity scattering causes the
decays of quasiparticle states and leads to the damping of spin excitaitons
associated with Im$\chi_{0}^{(ij)}({\bf Q},\omega)$. It gives rise to the
broadening of the peak. Exactly this bahavior is observed in Fig.1. Also
one can see from Fig.2 that the peak in Im$\chi_{0}^{(ij)}({\bf Q},\omega)$ 
disappears gradually upon the introduction of impurities. It is because the 
impurity scattering will wash out the van Hove singularity. Consequently,
no clear reasonance peak is observed at large impurity concentrations 
(e.g.$\Gamma/\Delta_{0}=0.08$) due to this effect and the damping of  
spin excitations.
Another feature in Fig.1 is that no significant excitation spectrum
weight has been found in the spin gap. The origin of
the spin gap in the SC state arises from a lack of thermally exciting 
$p-h$ pairs across the SC gap with transition wavevector ${\bf Q}$ when the
exciting energy is lower than the threshold $E_{th}\approx 2\Delta_{0}$. So,
though the impurity
self-energy produces an increase in the quasiparticle scattering rate, it may
be not strong enough to cause an observeable enhancement to the $p-h$
excitations across the SC gap. We note that the impurity vertex corrections 
entering Im$\chi_{0}^{(ij)}({\bf Q},\omega)$ consist of the $p-h$ ladder 
diagrams connected by the impurity scatteing lines, which may allow a strong 
scatterings and lead to singificant modification of the spin gap.

\section{VERTEX CORRECTION}
In the above calculations, the self-energy from the impurity scattering is 
considered to include the multiple scattering of quasiparticles from the
same impurity
in the noncrossing manner~\cite{quin}. Because the dynamical susceptibility
measured in magnetic neutron scattering is believed here to come from the
$p-h$ pair excitations, the multiple scattering of particles and holes from 
the same impurity should be examined. That is, we must include the vertex
corrections due to the impurity scattering, which is 
displayed diagrammatically in Fig.3. The single and double arrowed solid
lines in Fig.3 stand for the normal and pairing Green's functions of 
particles and holes renormalized by the impurity self-energy. The dashed 
line is the impurity interaction and the impurity is represented by a cross
${\bf\times}$. The multiple scatterings in the form of ladder diagrams and 
the multiple scatterings of quasiparticles from the same impurity can be 
explicitly seen from Fig.3 (b) and (c), respectively.
The vertex-corrected spin susceptibility can be written as a $4\times 4$ 
matrix equation~\cite{quin},
\begin{equation}
\hat{\chi}_{0}^{(ij)}({\bf q},i\omega_{m})=T\sum_{n}{\hat{M}^{(ij)}({\bf q},
i\omega_{m},i\omega_{n})
\over \hat{1}-I^{(ij)}(i\omega_{m},i\omega_{n})\hat{M}^{(ij)}({\bf q},
i\omega_{m},i\omega_{n})} ,
\end{equation}
where $\Gamma(i\omega_{m},i\omega_{n})=\hat{1}-I^{(ij)}(i\omega_{m},
i\omega_{n})\hat{M}^{(ij)}({\bf q},i\omega_{m},i\omega_{n})$ is the dressed
vertex and the impurity-scattering lines are given by,
\begin{equation}
I^{(ij)}(i\omega_{m},i\omega_{n})=-{n_{i} \over [\pi N_{0}]^{2}}
T^{(i)}_{0}(i\omega_{m}+i\omega_{n})T^{(j)}_{0}(i\omega_{n}) .
\end{equation}
The spin-triplet particle-particle channel into the $p-h$ bubbles by 
transfroming e.g. a spin-down particle into a spin-up hole and vice versa via
the mixing with the SC condensate is not included here, because its 
contributions to RPA normalized spin susceptibility is zero when one considers
the AF correlations in the form of that in $t-J$ model~\cite{brinck}. Thus, 
the components of $\hat{M}$ are,
\begin{eqnarray}
\hat{M}^{(ij)}_{11}({\bf q},i\omega_{m},i\omega_{n})
=&-\hat{M}^{(ij)}_{22}({\bf q},i\omega_{m}+2i\omega_{n},-i\omega_{n})
=-\hat{M}^{(ij)}_{33}({\bf q},-i\omega_{m}-2i\omega_{n},i\omega_{n})
\nonumber  \\
=&-\hat{M}^{(ij)}_{44}({\bf q},-i\omega_{m},-i\omega_{n})
=-\int {d^{2}p \over (2\pi)^{2}}G^{(i)}({\bf p+q},i\omega_{m}+i\omega_{n})
G^{(j)}({\bf q},i\omega_{n}) ,
\end{eqnarray}
\begin{equation}
\hat{M}^{(ij)}_{14}({\bf q},i\omega_{m},i\omega_{n})=
\hat{M}^{(ij)}_{23}({\bf q},i\omega_{m},i\omega_{n})=
-\int {d^{2}p \over (2\pi)^{2}}F^{(i)}({\bf p+q},i\omega_{m}+i\omega_{n})
F^{(j)}({\bf q},i\omega_{n}) ,
\end{equation}
where $G^{(i)}({\bf q},i\omega_{n})$ and $F^{(i)}({\bf q},i\omega_{n})$ are 
the normal and paring Green's functions of quasiparticles which has been 
renormalized by the impurity self-energy.

The equations (11), (13) and (14) are calculated by using the same method 
described in Sec.II. Results for Im$\chi^{ph}({\bf Q},\omega)$ are shown in 
Fig.4 for the same impurity concentrations as those in Fig.1. In contrast 
to the effect of self-energy corrections, an apparent contribution to spin 
excitation spectra is observed in the spin gap at large impurity concentrations
where no clear reasonance peak is observed. In order to separate the 
contribution of vertex corrections from self-energy corrections, we have 
carried out the similar calculations in which the Green's functions of the 
impurity-free system in $\hat{M}^{(ij)}$ are used. The result shows that the 
signal in the spin gap is solely due to the vertex corrections, meanwhile 
the magnitude of the spin gap as well as the position and the width of the 
peak remain unchanged, except for a slight enhancement of the peak height in 
the presence of only vertex corrections. The broad contribution in the spin 
gap may be understood as the
strong scattering involved in the impurity vertex which allows for the
multiple scatterings due to ladder diagrams. To address the reason
why this strong scattering takes effect mainly in low frequencies, we show 
in Fig.5 the decay rates of the quasiparticles due to the impurity 
self-energy implied by $1/\tau^{(i)}_{imp}(\omega)=-2{\rm Im}\Sigma^{(i)}_{0}
(\omega)=-2\Gamma T_{0}^{(i)}(\omega)$. The similar result has 
been obtained by Quinland and Scalapino~\cite{quin}. We can see that the 
decay rates increase as the frequency
decreases and reache its maximum at $\omega=0$. Because the impurity-scattering
lines in the vertex corrections is directly related to $\Sigma^{(i)}_{0}(\omega)$
as expressed in Eq.(12), this enhancement is amplified by the multiple 
scatterings in the form of ladder diagrams. We note that, in the absence of the 
vertex corrections, this enhancement is not strong enough to lead to an 
apparent spectral weight in the gap as discussed in Sec.II. From Fig.4, we 
can also see that the gaplike region in low frequencies still retains in the 
impurity-doping system . We may ascribe it to the fact
that the off-diagonal impurity self-energy $\Sigma_{1}$ vanishes identically
for a $d_{x^{2}-y^{2}}$ order parameter and therefore the angular (e.g.nodal)
structure of the SC gap is not changed. According to these features, we find 
that the overall modifications of the spin excitation spectra upon the 
doping of nonmagnetic impurity are in qualitatively consistent with the INS 
measurement on YBa$_{2}$(Cu$_{1-y}$Zn$_{y}$)$_{3}$O$_{6+y}$~\cite{sid,hara}. 
However, the spectral weight in the spin gap is still not large enough to 
account for quantitatively the experimental result~\cite{sid}. We note that 
a nonmagnetic impurity such as Zn in the CuO$_{2}$ planes is believed to
induce a local magnetic moment and lead to additional spin-flip
scattering~\cite{poil}. From the above discussion, we think that this
scattering may lead to more significant spectral weight in the spin gap than that
given here. Nevertheless, a detailed investigation of this effect is required
and will be carried out in future.

\section{CONCLUSION}
We have calculated the spin excitation spectra below $T_{c}$ for a model
$d_{x^{2}-y^{2}}$-wave superconductor with resonant impurity scattering. The
impurity self-energy corrections shift the position of the resonance peak to
low frequencies and broaden the peak. As the impurity concentration 
increases, the resonance peak disappears gradually. When no clear 
reasonance peak is observed, the impurity vertex corrections cause
a broad contribution to the excitation spectra in the spin gap, but the 
memory of the spin gap still retains. Thus, impurity-scatterings, together  
with the vertex corrections, account for qualitatively the experimental 
measurement on Zn-doping YBa$_{2}$Cu$_{3}$O$_{6+x}$.

\section*{ ACKNOWLEDGMENTS}
One of the authors (J.X.Li) acknowledge the support by National Nature
Science Foundation of China.

\newpage
\section*{FIGURE CAPTIONS}
\vspace{1cm}
Fig.1. Imaginary parts of the renormalized susceptibility
Im$\chi^{ph}({\bf Q},\omega)$ versus frequency $\omega$ for various impurity
concentrations $\Gamma/\Delta_{0}$. The solid line represents the result of
the pure system. Only the self-energy corrections are considered.

\vspace{0.5cm}
Fig.2. Imaginary Im$\chi_{0}^{(ba)}({\bf Q},\omega)$ and real 
Re$\chi_{0}^{(ba)}({\bf Q},\omega)$ parts of the bare susceptibility defined
in Eq.(7) versus frequency $\omega$ for various impurity concentrations
$\Gamma/\Delta_{0}$. 
$a$ and $b$ represent the antibonding and bonding bands respectively.
The result for $\chi_{0}^{(ab)}({\bf Q},\omega)$ is 
very similar to $\chi_{0}^{(ba)}({\bf Q},\omega)$.

\vspace{0.5cm}
Fig.3. Diagrammatic representation of the impurity vertex corrections to
the spin susceptibility in the dilute limit (see text).

\vspace{0.5cm}
Fig.4. Imaginary parts of the renormalized susceptibility
Im$\chi^{ph}({\bf Q},\omega)$ versus frequency $\omega$ for various impurity
concentrations $\Gamma/\Delta_{0}$. The solid line represents the result of
the pure system. Both the self-energy and vertex corrections are
considered.

\vspace{0.5cm}
Fig.5. Decay rates of quasiparticles in the antibonding band 
$1/\tau^{(a)}_{imp}(\omega)$ due to impurity self-energy corrections in the unitary limit. Results are shown for various impurity concentrations 
$\Gamma/\Delta_{0}$. The result for the bonding band is very similar to that shown here.
\end{document}